\documentclass[prc,aps,preprint,showpacs]{revtex4}
\usepackage{graphicx}

\newcommand{\simleq}{\; \raisebox{-0.4ex}{\tiny$\stackrel
{{\textstyle<}}{\sim}$}\;}
\begin{document}
\title
{Change of shell structure and magnetic moments 
of odd-N deformed nuclei towards neutron drip line} 

\author{ Ikuko Hamamoto$^{1,2}$ }

\address{
$^{1}$ {\it Division of Mathematical Physics, Lund Institute of Technology 
at the University of Lund, Lund, Sweden}   \\
$^{2}$ {\it The Niels Bohr Institute, Blegdamsvej 17, 
Copenhagen \O,
DK-2100, Denmark} \\ 
}




\begin{abstract}
Examples of the change of neutron shell-structure in both weakly-bound 
and resonant neutron
one-particle levels in nuclei towards the neutron drip line are exhibited.  
It is shown that the shell-structure change due to the weak binding 
may lead to the deformation of those nuclei 
with the neutron numbers 
$N \approx$ 8, 20, 28 and 40, which are known to be magic numbers 
in stable nuclei.
Nuclei in the ''island of inversion'' are most easily and in a simple manner
understood in terms of deformation.  As an example of spectroscopic properties
other than single-particle energies, magnetic moments of some weakly-bound 
possibly deformed odd-N nuclei 
with neutron numbers close to those traditional magic numbers are given, 
which are calculated using the wave
function of the last odd particle in deformed Woods-Saxon potentials.

\end{abstract}


\maketitle

\section{INTRODUCTION}
Some neutron-rich nuclei towards the neutron drip line, which have the neutron
numbers $N \approx$ 8 and 20, are now widely recognized to be deformed and 
often called nuclei in the ''island of inversion'' \cite{RKP69,CT75,EKW90}.    
In this paper the change of shell structure in weakly-bound and one-particle 
resonant neutron levels coming from a unique behaviour of small-$\ell$ 
neutron levels compared with large-$\ell$ levels is exhibited.  The 
behaviour comes from the strong $\ell$-dependence of the heights of 
centrifugal barrier, which are
proportional to $\ell (\ell +1) / R_{b}^{2}$ where $R_{b}$ is slightly 
larger than the 
nuclear radius.  The centrifugal barrier is absent for $s$ neutrons which can 
thus freely extend to the outside of nuclei as the binding energy approaches 
zero. 
Consequently, the energy of weakly-bound $s$ neutron levels is relatively 
insensitive to the strength of the potential.   In contrast, large-$\ell$
neutrons are confined to the inside of the potential and, thus, sensitive to 
the strength of the potential.
The barrier becomes increasingly higher both for larger $\ell$ neutrons and 
for smaller nuclei.  
As an example, in figure 1 we sketch the relative energies of 
three one-neutron levels in the $sd$ shell depending on binding
energies.  
The figure illustrates how the levels with smaller $\ell$-values shift
downwards relative to those with larger $\ell$-values, as the potential
strength becomes weaker. 
The cases of the strongly bound (such as by 30 MeV) levels and those 
bound by about 10 MeV 
(approximately equal to the Fermi level of stable $sd$ shell nuclei) 
can be found 
from figure 2-30 of \cite{BM69}, while the level order of very weakly bound
or one-particle resonant neutrons 
can be found for example in figure 1 of  \cite{HM03,IH07}.  
The level scheme being bound by about 10 MeV is similar to that of
the conventional modified-oscillator potential plus a spin-orbit potential. 
This similarity is accidental.  
Namely, this never means that "the one-particle level scheme of 
weakly-bound neutrons becomes similar to that of the harmonic oscillator 
potential (plus a spin-orbit potential)", which is unfortunately 
often stated in some publications.  Indeed, the harmonic oscillator potential 
has no surface and, consequently, using the potential 
it is impossible to express some phenomena 
which are unique in weakly-bound particles.

The similar change of the relative level order and the shell structure, 
which is sketched in figure 1 for 
the $sd$ shell, occurs for other shells.  For example, in the $pf$ shell 
the $p_{3/2}$ and $p_{1/2}$ levels are lowered relative to 
the $1f_{7/2}$ and $1f_{5/2}$ levels as the potential becomes weaker or the
binding energies become smaller, since the
centrifugal barrier for $\ell$=1 neutrons is six times lower than 
that for $\ell$=3 neutrons. 
This change of the shell structure 
can lead to the deformed ground state of some nuclei with the
traditional magic number for neutrons.
See examples in section 3.

Whether a given nucleus with weakly-bound neutrons will deform or not depends,
of course, also on the proton number, since some proton number may definitely
favor spherical shape.  For example, the proton number Z=8, oxygen isotope, 
has been known to prefer spherical shape.  Nevertheless, it was recently
reported \cite{HI09} that the unbound nucleus $^{12}_{8}$O$_{4}$ outside the
proton drip line is deformed in the way similar to its mirror nucleus 
$^{12}_{4}$Be$_{8}$. 

It is important to notice that the notion of one-particle states 
in deformed nuclei can be much more widely,
in a good approximation, applicable than that in spherical nuclei. 
This is because in the deformed mean field the major part of the long-range 
two-body interaction in the spherical mean field is 
already taken into account in the mean field.  Thus, the spectroscopy 
of deformed nuclei is often much simpler than that of spherical 
(vibrating) nuclei.  For example, we note that 
the analysis of observed spectroscopic 
properties of low-lying states in light mirror nuclei, 
$^{25}_{12}$Mg$_{13}$ and $^{25}_{13}$Al$_{12}$, 
in terms of one-particle motion in deformed (Nilsson) potential is very 
successful \cite{BM75}. 
As seen in those examples, 
the numerical results of the intrinsic configurations, which are obtained 
in the present work for the ground state of nuclei with the
neutron number of $N_{magic} \pm 1$, can be equally applicable to the low-lying
excited states (or isomeric states) of the neighboring nuclei.

As the one-particle energy $\varepsilon_{\Omega} < 0$ approaches zero 
in a deformed potential,
the probability of $\ell = 0$ component in the wave function approaches
unity in all $\Omega^{\pi}$=1/2$^{+}$ bound neutron orbits.  However, 
the energy,
at which the $s$-dominance shows up, depends on both deformation and 
individual orbits.   On the other hand, in the case of $\Omega^{\pi}$=1/2$^{-}$
and $\Omega^{\pi}$=3/2$^{-}$ the $p$-components increases as 
$\varepsilon_{\Omega} (<0) \rightarrow 0$, but the probability of $\ell = 1$
components at $\varepsilon_{\Omega} \Rightarrow$ 0 is less than unity 
and depends 
on individual levels and deformations. 

Except in very light nuclei a significant change of 
shell structure in weakly-bound
protons is not expected because of the presence of the Coulomb barrier.  
For this reason the shell structure and magnetic moments of proton-drip-line
nuclei are not included in this work.  On the other hand, 
the magnetic moment of
deformed odd-Z nuclei towards the neutron drip line is 
a very useful and interesting quantity, 
since it may clearly show possible deformation of those nuclei.  
However, the proton separation
energy of those neutron-rich nuclei is so large that the shell structure as well
as magnetic moments can be reasonably evaluated using some old traditional 
models, for example, the modified-oscillator model.
Therefore, the numerical calculations are not included here.

When possible rotational spectra or strongly-enhanced quadrupole moments, 
which are the direct sign of deformation, are difficult to
be measured in neutron drip line nuclei, an indication 
of deformation is to observe the unusually low-lying 2$^+$ state 
in even-even nuclei.  On the other hand, 
one-particle motion in the mean field that shows the shape of
nuclei can be easily recognized by studying the low-energy spectra of odd-A
nuclei \cite{BM75}.  In particular, the spin-parity of low-lying states or 
magnetic moments of the ground or 
isomeric states of odd-A drip-line nuclei 
may be sometimes more easily measured 
and can be used as a clear indication of possible deformation.
Therefore, in the present paper we survey the Nilsson diagrams that can be
applicable to weakly-bound nuclei with $N \approx$ 8, 20, 28 and 40, and give
the estimated values of magnetic moments of possibly deformed odd-N nuclei with
the neutron number close to those traditional magic numbers. 

In section 2 the essential points of the model used in the present work are
briefly summarized.  Numerical examples for nuclei in the $N \approx 8$ region 
is given in section 3.1, the $N \approx 20$ region in section 3.2, the 
$N \approx 28$ region in section 3.3, and the region $N \approx 40$ region in
section 3.4.  Conclusions and discussions are given in section 4.

\section{MODEL}
Recognizing the great usefulness of the Nilsson diagram, in which 
one-particle energies are plotted 
as a function of deformation parameter for a given potential, 
in the present article we apply the model and idea presented in 
\cite{IH07a}.  
The quadrupole deformation parameter $\beta$ is defined in \cite{IH04}. 
The coupled equations derived from the Schr\"{o}dinger equation are solved in
coordinate space with the correct asymptotic behaviour of wave functions for 
$r \rightarrow \infty$, both for bound \cite{IH04} and resonant 
\cite{IH05,IH06} levels.
In particular, one-particle resonant levels in a deformed potential are
estimated using the eigenphase formalism \cite{RGN66}.  Namely, one-particle
resonance is obtained if one of calculated eigenphases increases through 
$\pi /2$ as one-particle energy increases.  One-particle resonance is not
obtained if none of the calculated eigenphases increase through $\pi /2$ 
as energy increases.  The definition is a natural 
extension of the definition of one-particle resonance for spherical potentials 
in terms of phase shift, which can be found in standard textbooks.  
One-particle resonance with $\Omega ^{\pi}$=1/2$^+$ is not at all obtained 
if for $\varepsilon_{\Omega} (<0) \rightarrow 0$ 
the $\ell$=0 component of the wave function inside the potential exceeds 
a certain probability \cite{IH06}. 
On the other hand, for $\Omega^{\pi} \neq 1/2^{+}$ one-particle resonant
levels can be always found at least for small positive energies.
Compared with the Nilsson diagram based on modified oscillator potentials, the
striking difference of the level scheme exhibited in the present work comes 
from the behaviour of levels with low $\ell$ values (in particular, $\ell$=0 and
1) for $\beta$=0 and those with small $\Omega$ values 
(in particular, $\Omega^{\pi}$ = 
1/2$^+$, 1/2$^-$, and 3/2$^-$) for $\beta \neq$ 0, in both the weakly-bound 
and positive-energy regions. 

We use the parameters of Woods-Saxon potentials taken from the standard ones
\cite{BM69} for stable nuclei except the depth, $V_{WS}$.  
The potential depth for neutrons is adjusted so that the energy of the 
one-neutron level last occupied in a given odd-N nucleus is close to 
the measured neutron separation energy.  The diffuseness, the strength of 
the spin-orbit potential and the radius parameter
are, for simplicity, taken from those on p.239 of  \cite{BM69}.  
Considering the possible
contribution by weakly-bound neutron(s) to the self-consistent potential, 
a slightly larger diffuseness might be appropriate for presently studied nuclei.
However, it is noted that the major part of the nuclear potential is provided by
well-bound nucleons of the core.  Moreover, a larger diffuseness leads to the
degeneracy of the 2$s_{1/2}$ and 1$d_{5/2}$ (2$p_{3/2}$ and 1$f_{7/2}$) levels 
already at a larger binding energy than the one exhibited in the present
article. 
It is also remarked that for a given deformation the N-th 
deformed one-particle orbit filling-in all lower-lying Nilsson levels 
is almost uniquely determined for a given one-particle 
energy and within a reasonable variation of Woods-Saxon potential parameters.

The magnetic dipole moments of odd-N deformed nuclei are calculated using
equations
(4-86), (4-87), and (4-88) of  \cite{BM75}, while the matrix elements of 
$\ell_{\nu}$ and $s_{\nu}$ are evaluated using one-particle wave-functions 
in the deformed Woods-Saxon potential.  
See equations (5-86) and (5-87) of  \cite{BM75}.  The
tabulated magnetic moments are calculated for the band-head states for given
intrinsic configurations.  The spin of the band-head state is $I = \Omega$ for
$\Omega \neq 1/2$.  For $\Omega$=1/2 configurations the
decoupling parameter $a$ is calculated, and the spin of the band-head state is
different from I=1/2 for both $a < -1$ and $a > 4$.  See  \cite{BM75}.

The major part of the reduction of the effective $g_s$ factor, $g_s^{eff}$, 
from $g_s^{free}$ in the present model is supposed to come from the spin
polarization of deformed even-even core.  
For large deformation where the asymptotic
quantum numbers [N n$_z$ $\Lambda$ $\Omega$] become good quantum numbers, the
spin polarizations of $\Delta K$=0 type vanish, since the quantum numbers
$\Lambda$ and $\Sigma$ (= $\Omega$ $-$ $\Lambda$) become constants of the motion.
Consequently, the longitudinal $g_s^{eff}$ factor may approach $g_s^{free}$,
while the transverse matrix element ($g_K$ $-$ $g_R$)b in equation (4-86) of
\cite{BM75} is still affected by the
presence of $\Delta K$=1 polarizations of deformed even-even core.  
The latter
contributes only to the magnetic moments of K=1/2 bands.  The quenching of the
spin fluctuations in the large deformation limit seems to be only partially
achieved for the equilibrium deformations of stable deformed nuclei \cite{BR64}.
(A further reduction in the spin polarization of nuclei presently studied may
come from the fact that loosely-bound neutrons couple weakly with strongly-bound
core nucleons.)  In the present numerical calculations we use
$g_s^{eff}$=$g_s^{free}$ in the lightest mass ($N \approx 8)$ region where
deformed nuclei have larger deformation, while in heavier mass regions
$g_s^{eff}$=(0.7)$g_s^{free}$ is used.  We do not try any detailed evaluation 
of the spin polarization effect in individual deformed nuclei, since our present
aim is to show the spin-parity as well as magnetic moments for deformed shape of
nuclei studied, which are very different from those for spherical shape.  

In the present coupled channel calculation the $j$ = 1/2, 3/2, 5/2, 7/2, 9/2, 
11/2 and 13/2 channels 
are included for positive-parity levels, 
while the $j$ = 1/2, 3/2, 5/2, 7/2, 9/2 and 11/2 channels 
for negative-parity levels.

\section{NUMERICAL EXAMPLES}
\subsection{Odd-N neutron-rich nuclei in $N \approx 8$ region}
As it is known that the shape of the Be isotope can be very different 
for the A and A+1 nuclei, the notion of the mean field is not so well 
established in nuclei of this very light mass region.  
Nevertheless, in figure 2 the Nilsson
diagram is shown, of which the parameters are adjusted to the nucleus 
$^{17}_{6}$C$_{11}$, since in this mass region the spin-parity and/or 
magnetic moments 
are measured for some odd-N nuclei with weakly-bound neutron(s). 
In figure 2 the [200 1/2] level can be obtained as a one-particle resonant level 
as far as $d$ components are dominant.  In contrast, 
for $\beta > 0.46$ the resonant
level is not obtained because the $s$ component becomes dominant in the
one-particle wave-function.  
Similarly, due to the $s$-component dominance 
the extension of the [220 1/2] level to the
positive-energy region on the oblate side 
as a resonant level is not possible for $\beta < -0.12$.  

In figure 2, it is seen that in the spherical limit ($\beta$=0) the very weakly
bound $1d_{5/2}$ and $2s_{1/2}$ levels are almost degenerate.  Namely, as the
binding becomes very weak, the energy eigenvalue of the $s$ orbit 
shifts downward 
relative to that of the $d$ level, as schematically exhibited in figure 1.  
The calculated and measured magnetic moments of the ground state of 
$^{17}_{6}$C$_{11}$ and $^{11}_{4}$Be$_{7}$ are compared in table 1.
It is remarked that for a much weaker strength of the potential such as the one
applicable for $^{11}$Be, in which the [220 1/2] and [101 1/2] orbits at 
$\beta \approx 0.6$ are bound by several hundreds keV, the calculated bound
one-particle level scheme and magnetic moments around $\beta$=0.6 are found to 
remain nearly the same.  
When in a given region of deformation 
the curve of the Nilsson level is almost a straight line as a function of
deformation, the one-particle wave-function depends very little 
on deformation and, 
consequently, the calculated magnetic moment remains almost independent of
deformation.
We notice that measured magnetic moment of $^{15}$C$_{9}$ (I=1/2$^+$) that can
be a spherical nucleus, $|\mu| = 1.720 \pm 0.009$ $\mu_{N}$ \cite{KA02}, 
is nearly equal to that of $^{11}$Be$_{7}$  (I$^{\pi}$=1/2$^+$) 
that is most easily interpreted as a deformed nucleus, 
assuming that the sign of the former is minus.  
This is theoretically expected because the [220 1/2] wave-function of the 7th 
neutron in $^{11}$Be consists exclusively of $1d_{5/2}$ and $2s_{1/2}$
components, 
of which non-diagonal matrix element of magnetic dipole operator is zero, 
and, furthermore,  
the neutron one-particle magnetic moments of $1d_{5/2}$ and $2s_{1/2}$ 
are the same.

The spin-parity (3/2$^+$) of the ground state of $^{17}$C$_{11}$ is in a very
natural way interpreted as the band-head state of the N=11th neutron occupying
the [211 3/2] level.  The near degeneracy of the 1$d_{5/2}$ and 2$s_{1/2}$ 
levels at $\beta$=0 
can well be the reason why the nucleus $^{17}$C is deformed 
(Jahn-Teller effect).  The measured
magnetic moment is in good agreement with the calculated value based on the
deformation, as seen in table 1.  

\subsection{Odd-N neutron-rich nuclei in $N \approx 20$ region}
In figure 3 the Nilsson diagram is shown, of which the parameters are 
approximately adjusted to
the nucleus $^{31}_{10}$Ne$_{21}$.  
For a slightly stronger potential which may be applicable for 
$^{31}_{12}$Mg$_{19}$ (the neutron separation energy $S_n$ = 2.38 MeV) 
and $^{33}_{12}$Mg$_{21}$ 
($S_n$ = 2.22 MeV), the calculated bound one-particle level scheme and
magnetic moments remain nearly the same as those presented here.  
See figure 3 of  \cite{IH07a}.   
Neither the $2p_{3/2}$ nor $2p_{1/2}$ levels are obtained as one-particle
resonant levels for the present potential.    
Nevertheless, in figure 3 
the approximate positions of those levels, 
which are extrapolated from the resonance energies 
obtained for slightly stronger 
spherical Woods-Saxon potentials, 
are indicated in figure 3 with the question mark.
It is noted that when both the $2p_{3/2}$ and $1f_{7/2}$ levels appear as very 
low-lying resonant levels, the $2p_{3/2}$ level may lie lower than the
$1f_{7/2}$ level.
 
The second lowest $\Omega^{\pi}$=1/2$^-$ level for $\beta > 0$ 
denoted by the dotted curve in figure 3 cannot continue 
for $\beta < 0.33$ as a
one-particle resonant level, 
because one-particle resonant levels with the major
component of $\ell$=1 cannot survive for higher $\varepsilon_{\Omega}$ values. 
The complicated behaviour of the resonant level expressed by the dotted curve 
for $\beta < 0$ indicates clearly the influence 
of the $2p_{3/2}$ and $2p_{1/2}$ levels that do not explicitly 
appear in figure 3. 
It is noted that for a pure $1f_{7/2}$ shell the dotted curve increases almost
linearly as $| \beta |$ increases for $\beta < 0$, just as a smooth 
extension of 
the curve for $\beta > 0$.   
Around $\beta = -0.4$ the major component of the $\Omega^{\pi}$=1/2$^-$ level 
is $\ell$=1.  
As $\beta (<0)$ increases $\varepsilon_{\Omega}$ increases, and the
one-particle resonant level with the $\ell$=1 major component can hardly 
survive for $\varepsilon_{\Omega} > 1.2$ MeV, where both the width and 
$\varepsilon_{\Omega}$ rapidly increase. The major component of the 
one-particle resonant level expressed by the dotted
curve for $-0.2 < \beta < 0.07$ is $\ell$=3, and consequently the resonant 
level is well defined with the reasonably small width.
 
The near degeneracy of $1f_{7/2}$, $2p_{3/2}$ and $2p_{1/2}$ levels 
at $\beta$=0
can be the origin of possible deformation that appears 
in the system with a few weakly bound neutrons
occupying the $1f_{7/2}$-$2p_{3/2}$-$2p_{1/2}$ shell (Jahn-Teller effect). 
Namely, using the degeneracy the energy of a particular combination of 
the $pf$ components can be made lower when deformation sets in.  

In this region of ''island of inversion'' there are two odd-N nuclei of which
magnetic moments of the ground state are measured. 
The measured magnetic moment \cite{GN05} of the ground state 
(I$^{\pi}$=1/2$^+$) of $^{31}_{12}$Mg$_{19}$ 
is $-$0.88355(15) $\mu_{N}$, while it is reported in  \cite{DY07} that 
the magnetic moment of $^{33}_{12}$Mg$_{21}$ is $-$0.7456(5) $\mu_{N}$ and 
the spin is 3/2.  Though the parity of the ground state of $^{33}$Mg is
currently still under debate (for example, see the recent $\beta$-decay study 
in \cite{VT08}), from table 2 the measured negative magnetic moment seems to
have no other choice than negative parity for the ground state. 
It is remarked that the Nilsson orbit that can be occupied by the 21st neutron
for $0 < \beta \simleq 0.6$ is either [330 1/2] or [202 3/2] or [321 3/2] within
the reasonable variation of potential parameters.  All of the three Nilsson
orbits lead to I=3/2 for the band-head spin.  
Considering that we have fixed parameters $g_s^{eff}$ and $g_R$ in the numerical
calculation of magnetic moments, of which 
some standard values are used, the agreement of the calculated 
and measured values in table 2 is very good. 
In table 2 calculated magnetic moments are tabulated for a given $\beta$ value,
however, it should be noted that calculated values remain approximately
constant, as far as the Nilsson level is almost a straight line as a function 
of $\beta$.  

\subsection{Odd-N neutron-rich nuclei in $N \approx 28$ region}
Odd-N nuclei with weakly bound neutrons in this region are currently under
study and, to our knowledge, no measurements of magnetic moments 
are reported. 
Examining E(2$_1^+$) of even-even nuclei with N=28, possible candidates for
the deformed ground state are $^{42}_{14}$Si$_{28}$ and $^{44}_{16}$S$_{28}$, 
of which E($2_1^+$) is 770 (19) \cite{BB07} and 1297(18) keV \cite{TG97}, 
respectively.  There have been both some shell-model calculation and 
more elaborated many-body calculation, which suggested an oblate deformation 
for the nucleus $^{42}$Si.  Thus, in this region we show magnetic 
moments of odd-N nuclei evaluated for both prolate and oblate shape.  
Though $^{40}$Mg lies inside the neutron drip
line, Mg isotope is not discussed here,  
since none of the neighboring odd-N nuclei, $^{41}$Mg and $^{39}$Mg, is bound.

In figure 4 we show the Nilsson diagram, of which the parameters are 
approximately adjusted to
$^{41}_{14}$Si$_{27}$.  The reported $S_n$ value \cite{BJ07} of this nucleus 
is 1.34 MeV with a rather large ambiguity.  It is seen that at $\beta$=0 
the $2p_{3/2}$ level lies only 1.2 MeV higher than the $1f_{7/2}$ level 
when the former is bound only by 0.28 MeV.  Then, there may be a good chance 
for N=27 nuclei being deformed, since the neutron number N=28 may not work 
as a magic number.

Calculated magnetic moments are shown in table 3.  It is interesting to see 
that the magnetic moment of $^{41}$Si can be very different if it has an oblate
shape. This is because the (positive or) almost vanishing value of the
calculated magnetic moment comes from a subtle balance between the p$_{3/2}$
and p$_{1/2}$ components in the one-particle wave-function of the [301 1/2]
orbit.  That means, the calculated magnetic moments of oblate nuclei with N=27 
depend on the size of the oblate deformation, as can be guessed 
from the curved (and not straight-line) Nilsson level in figure 4 
as a function of $\beta$.

In the $g$-factor measurement of $^{43}_{16}$S$_{27}$, which is an isotone of 
$^{41}$Si, it is reported \cite{LG09} 
that the 320 keV isomeric state has $7/2^-$ and $g$=$-$0.317(4).  
This observation of the excited state indicates that 
the ground state of $^{43}$S is deformed.

In figure 5 the Nilsson diagram for neutrons is shown, 
of which the parameters are approximately adjusted 
to $^{45}_{16}$S$_{29}$.  
The comparison between figure 4 and figure 5 clearly shows that the N=28 shell-gap
at $\beta$=0, namely the distance between the $2p_{3/2}$ and $1f_{7/2}$ levels,
becomes increasingly smaller as the binding energy of the $2p_{3/2}$ level
approaches zero.  This appreciable change of the N=28 
shell-gap indicates that for 
a given neutron-number N$\approx$28 the Si isotope has 
a better chance to be deformed than the S isotope.  
The nucleus $^{43}_{14}$Si$_{29}$ is known to lie inside the
neutron drip line, however, to our knowledge 
no spectroscopic information is yet available.
In figure 5 the asymptotic quantum numbers, [N n$_z$ $\Lambda$ $\Omega$], are
written on the Nilsson levels which may be occupied by the N=29th neutron, 
while
calculated magnetic moments, in which the last odd neutron is placed in
respective Nilsson orbits, are given in table 4. 

\subsection{Odd-N neutron-rich nuclei in $N \approx 40$ region}
Neutron-drip-line nuclei in the region of N=40 are still far away from
reaching experimentally.  Moreover, N=40 is not one of magic numbers in the j-j
coupling shell model.  Nevertheless, we include this subsection, because; 
(a) while there are no deformed N=40 nuclei along the stability line, recent 
experimental studies show relatively low energies ($\approx$ 500 keV) 
of the first excited 2+ state 
of even-even nuclei of both $_{26}$Fe and $_{24}$Cr isotopes 
in the region of N $\approx$ 40.  
Indeed, 
it is stated \cite{PA08} that the nucleus $^{64}_{24}$Cr$_{40}$ may be a center
of deformation in this region; (b) Some magnetic-moment measurements of
isomeric states, the presence of which is not rare in this region, 
have already been reported. 

Odd-N nuclei of Fe and Cr isotopes with N$\approx$40 
around $^{64}_{24}$Cr$_{40}$ have measured $S_n$ values of a few MeV. 
In figure 6 the Nilsson diagram is shown, which may be
useful for the Fe and Cr isotopes with N$\approx$40.  
It is noted that the $2p_{1/2}$ and $1f_{5/2}$ levels in figure 6 
are nearly degenerate
in contrast to the level scheme shown in figures 4 and 5. This is because 
in the weaker potentials of the latter figures 
the $2p_{1/2}$ level becomes increasingly lower
relative to the $1f_{5/2}$ level.  
Calculated magnetic moments
are given in table 5, in which those given for $\beta$=0.25 remain nearly the
same for $\beta$=0.35.  It may be speculated that the observed 
387 keV isomeric state in 
$^{67}_{26}$Fe$_{41}$ \cite{MS03} 
may be the $I^{\pi}$=1/2$^{-}$ or 5/2$^{-}$ state coming from the 
[301 1/2] or [303 5/2] intrinsic configuration if the isomeric state 
is deformed.

\section{CONCLUSIONS AND DISCUSSIONS}

Examples of the change of neutron shell-structure for weakly
bound neutrons are illustrated for neutron-rich nuclei with N$\approx$8, 20,
28 and 40.  Both weakly-bound and resonant one-particle levels are properly
calculated by directly solving the Schr\"{o}dinger equation in mesh of space
coordinate with the appropriate boundary condition.  Magnetic moments of
possibly deformed odd-N nuclei in the region are calculated using one-particle
wave-functions in deformed Woods-Saxon potentials.  

Among the examples taken in the present paper, the near degeneracy of the
weakly-bound $1d_{5/2}$ and $2s_{1/2}$ levels (figure 2), that of the one-particle
resonant $1f_{7/2}$, $2p_{3/2}$ and $2p_{1/2}$ levels (figure 3), and that of the
$1f_{5/2}$ and $2p_{1/2}$ levels bound by several MeV (figure 6) are 
the most clear-cut
examples of the unique shell-structure, which is very different from what is
known for traditional stable nuclei.  It is hoped that more experiments will 
pin down the unique shell structure.  In the present paper we have argued that 
the shell structure may lead to, among
others; (a) the appearance of the N=16 magic number \cite{AO00}; (b) possible
deformed shapes of $^{11-14}$Be \cite{HS07}, $^{17}$C, and $^{19}$C; (c) 
deformed nuclei in the island of inversion around N $\approx$ 20; (d) The
disappearance of the magic number N=28 and possible deformation of some
neutron-drip-line nuclei with N$\approx$28.

Calculated magnetic moments of the deformed ground states of $^{11}$Be, 
$^{17}$C, $^{31}$Mg, and $^{33}$Mg are in good agreement with already measured
ones, while the values of $g_s$ and $g_R$ have not been adjusted to individual
nuclei.  Except the case of $^{11}$Be where the unique magnetic moments of the
d$_{5/2}$-s$_{1/2}$ mixed neutron wave-functions play a role, 
both measured and calculated values of magnetic moment are
significantly different from what is expected from spherical shape and, thus,
can be used as a good identification of deformation.  
The measurement of magnetic moments of other neutron-drip-line nuclei is
strongly wanted.

Magnetic moments estimated in the present work are applicable not only to 
the ground state of the odd-N nuclei studied but also to 
excited or isomeric states in the neighboring nuclei.  The presence of 
isomeric states is often expected in the region of N$\approx$28 and N$\approx$40
neutron-rich nuclei due to  
the coexistence of spherical and deformed shape or the unique shell-structure.
Magnetic moments presently estimated for a given deformation remain nearly 
the same
when one-particle energies in the Nilsson diagram 
vary almost linearly as a
function of deformation, because the structure of 
single-particle wave-functions in those cases remains nearly independent of
deformation.

It should be remarked that the change of the shell structure 
described in the present paper is different from (and independent of) 
the one coming from the neutron-proton tensor force \cite{TO05}, which is 
currently very fashionable for being an origin of shell-structure change.  
The latter depends on the proton number of respective nuclei 
when the neutron shell-structure is discussed.
Moreover, the effect of the tensor force is usually estimated using harmonic
oscillator wave-functions in the shell model and, thus, the effect of 
weakly binding is not included.

The author is grateful to Dr. H. Ueno for introducing measured magnetic
moments of neutron-rich nuclei in the presently-studied mass region to her.

\newpage
\begin{table}[hbt]
\begin{center}

\caption{ 
Calculated magnetic dipole moments of the ground states of 
very light odd-N nuclei with one
weakly-bound neutron, in comparison with observed ones.  Measured neutron
separation energies are expressed by S$_n$.  Values of 
$g_{R}$=0.35 (for $\beta \neq$0) and
$g_{s}^{eff}$=$g_{s}^{free}$ are used.  Corresponding one-particle levels in
figure 2 are those labeled as [211 3/2], [220 1/2] and 2$s_{1/2}$ for 
$^{17}$C, $^{11}$Be and $^{15}$C, respectively.
}
\vspace{5pt}

\begin{tabular}{cccccc} \hline
Nucleus & S$_n$ & $I^{\pi}$ & $\mu _{obs}$ & Reference & 
$\mu_{calc}$ (at $\beta$, [N n$_z$ $\Lambda$ $\Omega$]) \\
& (keV) &  & ($\mu_{N}$) & & ($\mu_{N}$)  \\ \hline
$^{17}$C$_{11}$ & 727 & 3/2$^+$ & $\pm$0.758(4) & \cite{HU04} &
$-$0.75 ($\beta$=0.4, [211 3/2]) \\
$^{11}$Be$_{7}$ & 504 & 1/2$^+$ & $-$1.6816(8) & \cite{WG99} &
$-$1.7 ($\beta$=0.6, [220 1/2])  \\
$^{15}$C$_{9}$ & 1218 & 1/2$^+$ & $\pm$1.720(9) & \cite{KA02} &
$-$1.9 ($\beta$=0, 2$s_{1/2}$)  \\  \hline

\end{tabular}
\end{center}
\end{table}

\vspace{1cm}

\begin{table}[hbt]
\begin{center}

\caption{
Calculated magnetic dipole moments of N=21 and 19 nuclei, 
in which the last odd neutron is placed in corresponding Nilsson orbits 
for prolate shape, 
in comparison with available experimental data.  See the Nilsson diagram 
in figure 3. Values of $g_{R}$=0.38 and
$g_{s}^{eff}$ = (0.7)$g_{s}^{free}$ are used. 
Note that for spherical shape one obtains  
$\mu_{calc}(f_{7/2})$ = $-$1.3 $\mu_N$ and 
$\mu_{calc}(d_{3/2})$ = +0.80 $\mu_N$, respectively, 
using $g_{s}^{eff}$ = (0.7)$g_{s}^{free}$.
}
\vspace{5pt}

\begin{tabular}{ccccccc} \hline 
Nucleus & S$_n$ & $(I^{\pi})_{obs}$ & $\mu _{obs}$ & Reference & 
$(I^{\pi})_{cal}$ & 
$\mu_{calc}$ (at $\beta$, [N n$_z$ $\Lambda$ $\Omega$]) \\
& (keV) & & ($\mu_{N}$) & & & ($\mu_{N}$)  \\ \hline
$^{33}$Mg$_{21}$ & 2222 & 3/2 & $-$0.7456(5) & \cite{DY07} & 3/2$^-$ &
$-$0.88 ($\beta$=0.25, [330 1/2]) \\
& & & & & 3/2$^+$ & +0.91 ($\beta$=0.35, [202 3/2]) \\
& & & & & 3/2$^-$ & $-$0.39 ($\beta$=0.45, [321 3/2]) \\
$^{31}$Mg$_{19}$ & 2378 & 1/2$^+$ & $-$0.88355(15) & \cite{GN05} & 1/2$^+$ & 
$-$1.00 ($\beta$=0.45, [200 1/2])  \\ 
& & & & & 3/2$^-$ & $-$0.91 ($\beta$=0.35, [330 1/2]) \\  \hline

\end{tabular}
\end{center}
\end{table}

\vspace{1cm}

\begin{table}[hbt]
\begin{center}

\caption{
Calculated magnetic dipole moments of N=27 nuclei, 
in which the last odd neutron is placed in corresponding Nilsson orbits. 
See the Nilsson diagram in figure 4. 
Values of $g_{R}$=0.38 and $g_{s}^{eff}$ = (0.7)$g_{s}^{free}$ are used. 
Note that for spherical shape one obtains  
$\mu_{calc}(p_{1/2})$ = +0.4 $\mu_N$, $\mu_{calc}(p_{3/2})$ = $-$1.3 $\mu_N$ 
and $\mu_{calc}(f_{7/2})$ = $-$1.3 $\mu_N$, 
respectively, using $g_{s}^{eff}$ = (0.7)$g_{s}^{free}$.
}

\vspace{5pt}

\begin{tabular}{cccc} \hline 
Nucleus & S$_n$ &  $(I^{\pi})_{cal}$ & 
$\mu_{calc}$ (at $\beta$, [N n$_z$ $\Lambda$ $\Omega$]) \\
& (MeV) & & ($\mu_{N}$)   \\ \hline
$^{41}$Si$_{27}$ & 1.34 & 3/2$^-$ & +0.07 ($\beta$=$-$0.4, [301 1/2]) \\
& & 3/2$^-$ & $-$0.66 ($\beta$=0.25, [321 1/2]) \\
& & 5/2$^-$ & $-$0.58 ($\beta$=0.45, [312 5/2]) \\ \hline 

\end{tabular}
\end{center}
\end{table}

\vspace{1cm}

\begin{table}[hbt]
\begin{center}

\caption{
Calculated magnetic dipole moments of N=29 nuclei, 
in which the last odd neutron is placed in corresponding Nilsson orbits. 
See the Nilsson diagram in figure 5. 
Values of $g_{R}$=0.38 and $g_{s}^{eff}$ = (0.7)$g_{s}^{free}$ are used. 
Note that for spherical shape one obtains  
$\mu_{calc}(p_{1/2})$ = +0.4 $\mu_N$, $\mu_{calc}(p_{3/2})$ = $-$1.3 $\mu_N$ 
and $\mu_{calc}(f_{7/2})$ = $-$1.3 $\mu_N$, respectively, 
using $g_{s}^{eff}$ = (0.7)$g_{s}^{free}$.
}
\vspace{5pt}
\begin{tabular}{cccc} \hline 

Nucleus & S$_n$ &  $(I^{\pi})_{cal}$ & 
$\mu_{calc}$ (at $\beta$, [N n$_z$ $\Lambda$ $\Omega$]) \\
& (MeV) & & ($\mu_{N}$)   \\ \hline
$^{45}$S$_{29}$ & 2.21 & 7/2$^-$ & $-$0.74 ($\beta =$ 0.25, [303 7/2]) \\
& & 1/2$^-$ & +0.59 ($\beta$=0.45, [310 1/2]) \\
& & 1/2$^-$ & +0.59 ($\beta$=$-$0.40, [310 1/2]) \\ 
& & 3/2$^-$ & +0.16 ($\beta$=$-$0.40, [312  3/2]) \\  \hline

\end{tabular}
\end{center}
\end{table}

\vspace{1cm}





\begin{table}[hbt]
\begin{center}

\caption{
Calculated magnetic dipole moments of N $\approx$ 40 nuclei, 
in which the last odd neutron is placed in corresponding Nilsson orbits. 
See the Nilsson diagram in figure 6. 
Values of $g_{R}$=0.38 and $g_{s}^{eff}$ = (0.7)$g_{s}^{free}$ are used. 
Note that for spherical shape one obtains  
$\mu_{calc}(p_{1/2})$ = +0.4 $\mu_N$, $\mu_{calc}(f_{5/2})$ = +1.0 $\mu_N$ 
and $\mu_{calc}(g_{9/2})$ = $-$1.3 $\mu_N$, respectively, 
using $g_{s}^{eff}$ = (0.7)$g_{s}^{free}$.
}
\vspace{5pt}

\begin{tabular}{cccc} \hline 

Nilsson orbits & $\beta$ &  $(I^{\pi})_{cal}$ & $\mu_{calc}$ \\
([N n$_z$ $\Lambda$ $\Omega$]) & & & ($\mu_{N}$)   \\ \hline
([303 5/2]) & 0.25 & 5/2$^-$ & +1.1 \\
([301 1/2]) & 0.25 & 1/2$^-$ & +0.43 \\
([431 3/2]) & 0.25 & 3/2$^+$ & $-$0.19 \\
([422 5/2]) & 0.35 & 5/2$^+$ & $-$0.46 \\
([301 3/2]) & 0.35 & 3/2$^-$ & $-$0.40 \\  \hline

\end{tabular}
\end{center}
\end{table}

\vspace{3cm}

\mbox{ } 
 
\pagebreak[4]

\newpage

\noindent
\begin{figure}
\begin{center}
\includegraphics[width=12cm]{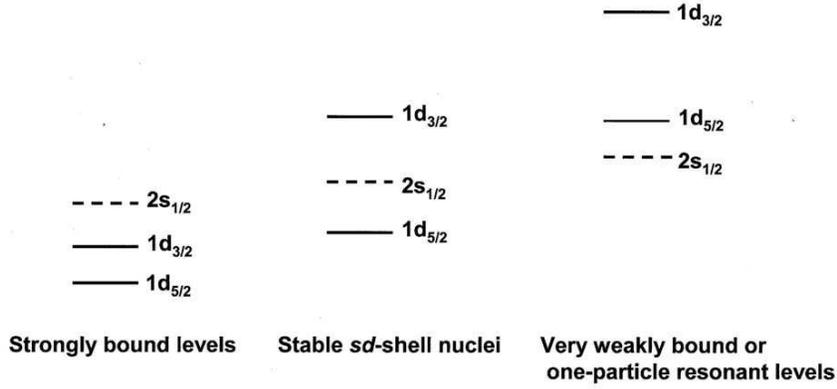}
\vspace{1cm}
\caption{
Qualitative sketch of relative energies of three one-neutron levels in the $sd$
shell depending on the binding energy or the potential strength. 
The level order sketched in the middle 
figure is the one known from stable $sd$ shell nuclei.}  
\label{fig1}
\end{center}

\end{figure}

\noindent
\begin{figure}
\begin{center}
\includegraphics[width=12cm]{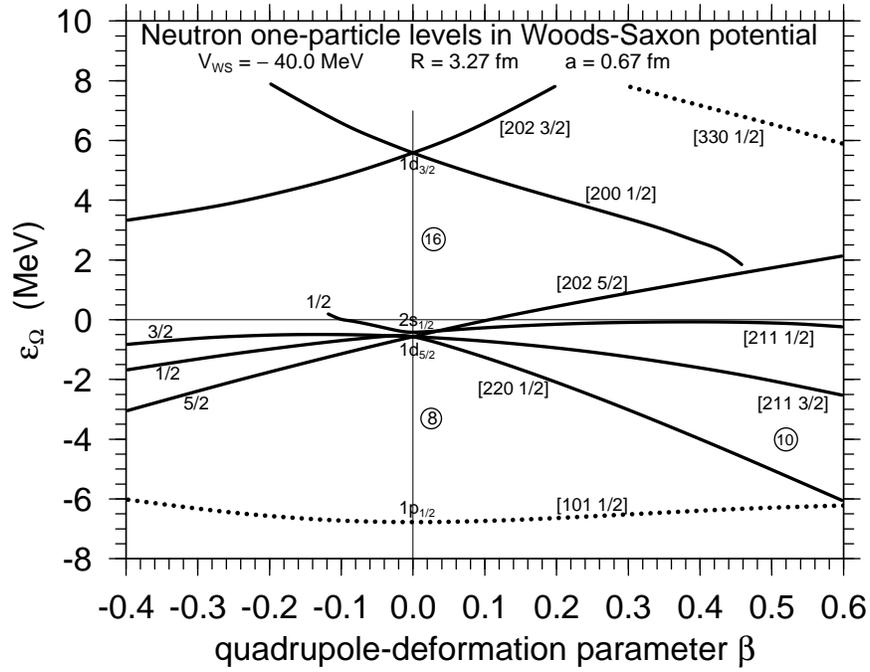}
\vspace{10cm}
\caption{
Neutron one-particle levels as a function of axially-symmetric 
quadrupole deformation.
Parameters of the Woods-Saxon potential are designed approximately for 
the nucleus $^{17}_{6}$C$_{11}$, of which $S_n$ = 730 keV.
The diffuseness, the radius and the depth of the Woods-Saxon potential are 
0.67 fm, 3.27 fm, and $-$40.0 MeV, respectively.
One-particle levels are denoted by the asymptotic quantum numbers, 
[$N n_z \Lambda \Omega$], on the prolate side ($\beta > 0$), while on the oblate
side ($\beta < 0$) the quantum number $\Omega$ is shown only in the case that it
is difficult to see as a continuation of the curve on the prolate side.
The $\Omega^{\pi}$ = 1/2$^{-}$ levels are denoted by dotted curves, 
while positive-parity levels are 
plotted by solid curves.
The neutron numbers 8, 10 and 16, which are obtained by filling in all 
lower-lying levels, are indicated with circles.
See the text for details.}
\label{fig2}
\end{center}

\end{figure}

\noindent
\begin{figure}
\begin{center}
\includegraphics[width=12cm]{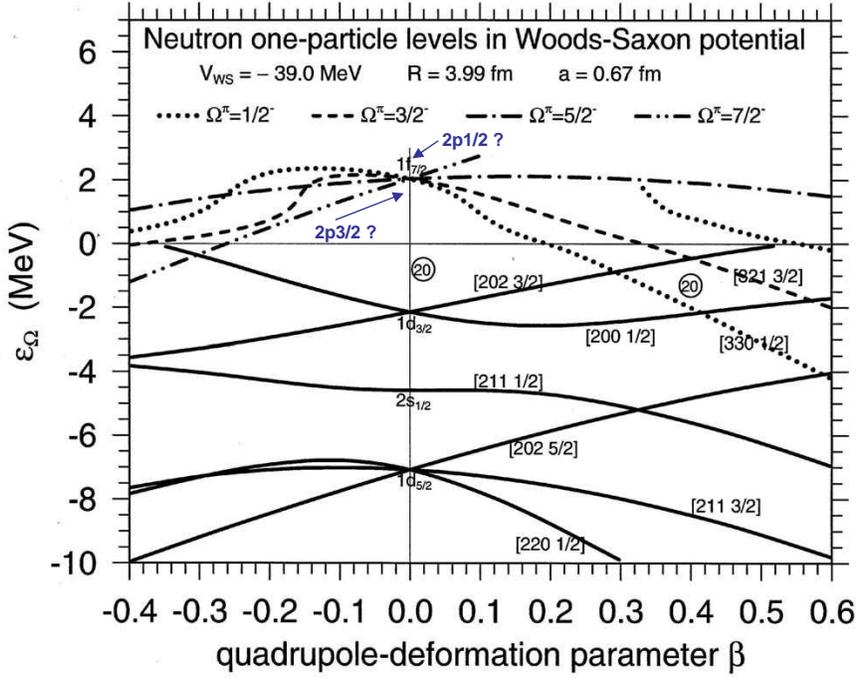}
\vspace{1cm}
\caption{Neutron one-particle levels as a function of axially-symmetric 
quadrupole deformation.
Parameters of the Woods-Saxon potential are designed approximately for 
the nucleus $^{31}_{10}$Ne$_{21}$, of which the measured $S_n$ value is 
(0.3$\pm$1.6) MeV.
The diffuseness, the radius and the depth of the Woods-Saxon potential are 
0.67 fm, 3.99 fm, and $-$39.0 MeV, respectively.
Some one-particle levels are denoted by the asymptotic quantum numbers, 
[$N n_z \Lambda \Omega$].
Positive-parity levels are plotted by solid curves.
The neutron number 20, which is obtained by filling in all 
lower-lying levels, is indicated with a circle. 
The 2$p_{3/2}$ and 2$p_{1/2}$ neutron levels at $\beta$=0 are not
obtained as one-particle resonant levels, 
and the next low-lying one-particle resonant 
level at $\beta$=0 is the $1f_{5/2}$ level found at 9.50 MeV.
The approximate positions of the 2$p_{3/2}$ and 2$p_{1/2}$ levels at $\beta$=0 
are indicated with ''?'', which are extrapolated 
from the resonant energies obtained by using a slightly stronger 
Woods-Saxon potential.   
See the text for details.}
\label{fig3}
\end{center}

\end{figure}

\noindent
\begin{figure}
\begin{center}
\includegraphics[width=12cm]{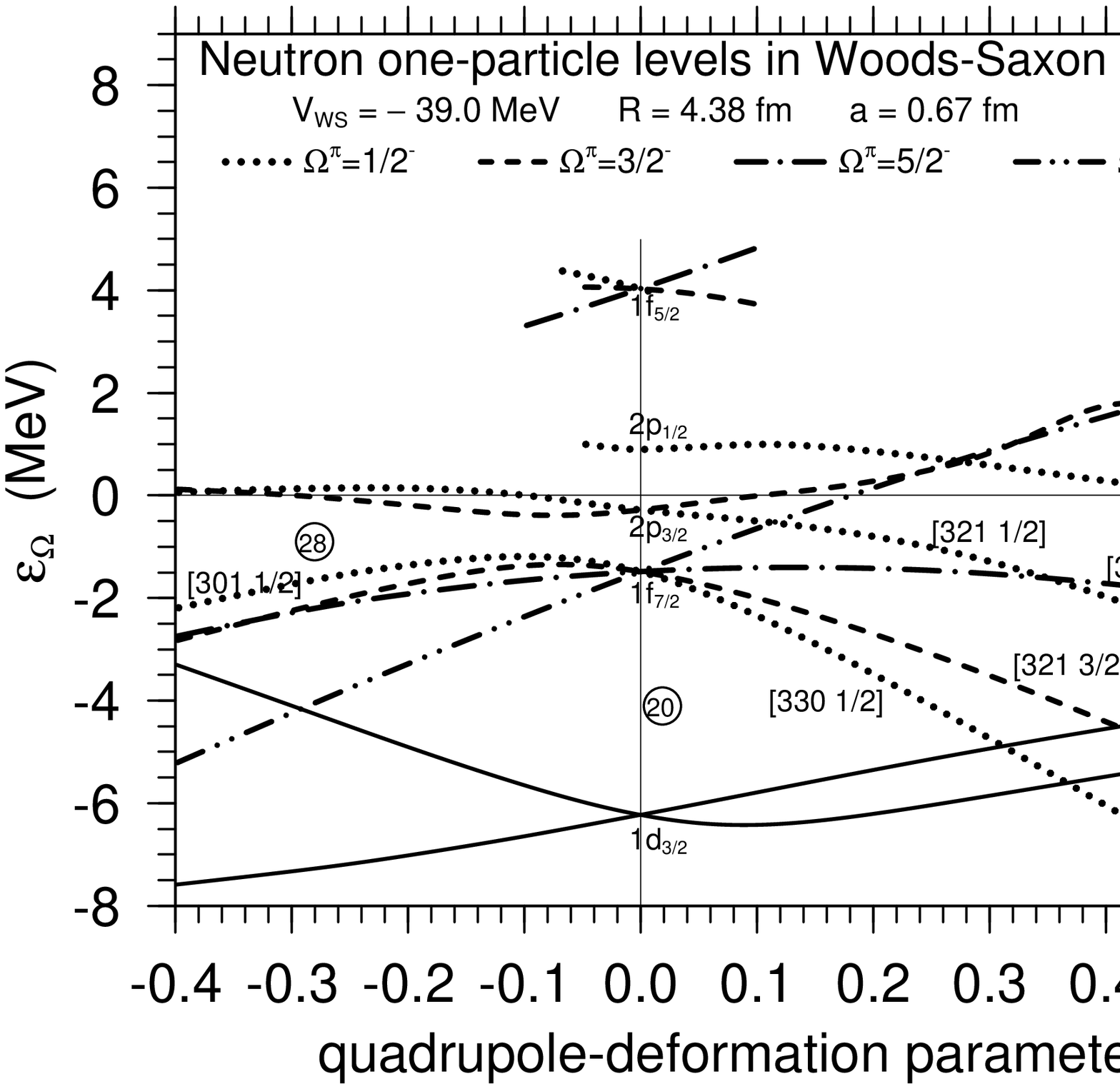}
\vspace{10cm}
\caption{Neutron one-particle levels as a function of axially-symmetric 
quadrupole deformation.
Parameters of the Woods-Saxon potential are designed approximately for 
the nucleus $^{41}_{14}$Si$_{27}$, of which $S_n$ = 1.34 MeV. 
The diffuseness, the radius and the depth of the Woods-Saxon potential are 
0.67 fm, 4.38 fm, and $-$39.0 MeV, respectively.
Some one-particle levels are denoted by the asymptotic quantum numbers, 
[$N n_z \Lambda \Omega$].
Positive-parity levels are plotted by solid curves.
The neutron numbers 20 and 28, which are obtained by filling in all 
lower-lying levels, are indicated with circles. 
The $\Omega^{\pi}$=1/2$^-$ level originating from 2$p_{1/2}$ at $\beta$=0 
does not survive as
one-particle resonance for $\varepsilon_{\Omega} > 1$ MeV and $\beta < -$0.05. 
On the other hand, for simplicity of the figure, 
only in the neighborhood of $\beta$=0 we have plotted 
resonant levels originating from $f_{5/2}$. 
See the text for details.}
\label{fig4}
\end{center}

\end{figure}

\noindent
\begin{figure}
\begin{center}
\includegraphics[width=12cm]{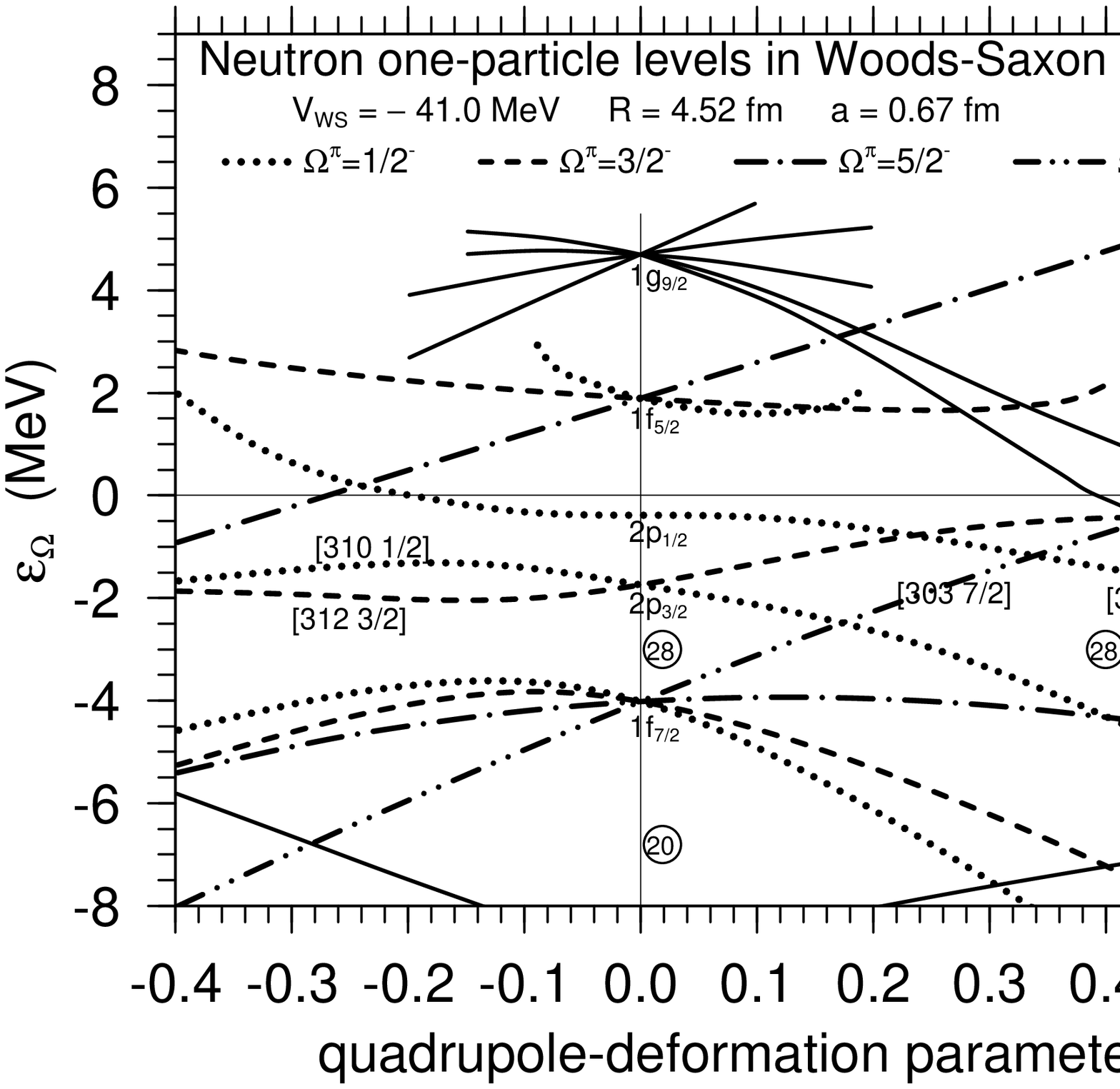}
\vspace{10cm}
\caption{Neutron one-particle levels as a function of axially-symmetric 
quadrupole deformation.
Parameters of the Woods-Saxon potential are designed approximately for 
the nucleus $^{45}_{16}$S$_{29}$, of which $S_n$ = 2.21 MeV
The diffuseness, the radius and the depth of the Woods-Saxon potential are 
0.67 fm, 4.52 fm, and $-$41.0 MeV, respectively.
Some one-particle levels possibly occupied by the N=29th neutron 
are denoted by the asymptotic quantum numbers, 
[$N n_z \Lambda \Omega$].
Positive-parity levels are plotted by solid curves.
The neutron numbers 20 and 28, which are obtained by filling in all 
lower-lying levels, are indicated with circles. 
Neutron resonant levels originating from 1$f_{5/2}$ at $\beta$=0 are plotted 
as far as they are obtained following the definition of the eigenphase
formalism.  On the other hand, not all neutron resonant levels originating from
1$g_{9/2}$ are plotted, for simplicity of the figure. 
See the text for details.}
\label{fig5}
\end{center}

\end{figure}

\noindent
\begin{figure}
\begin{center}
\includegraphics[width=12cm]{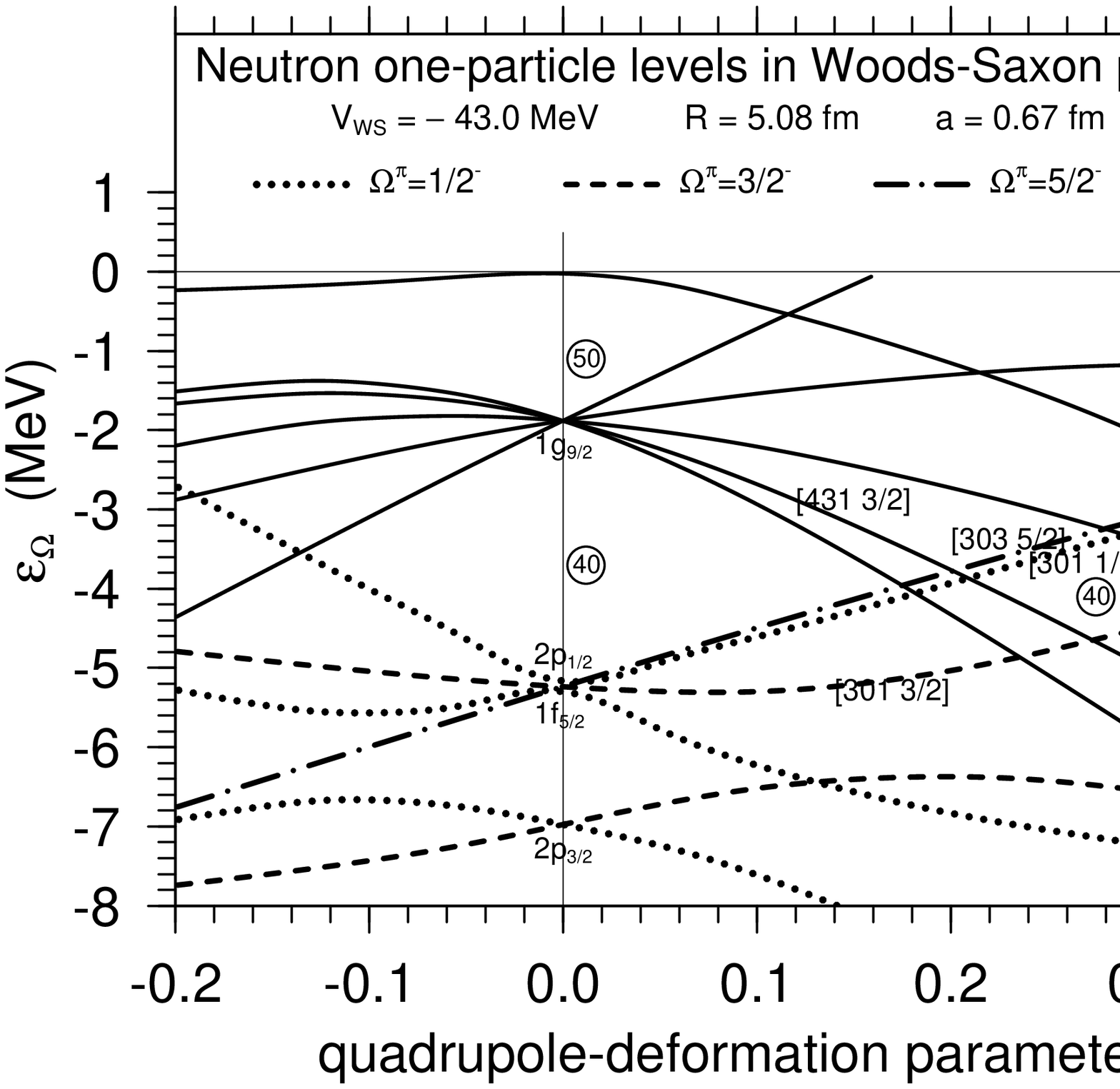}
\vspace{10cm}
\caption{Neutron one-particle levels as a function of axially-symmetric 
quadrupole deformation.
Parameters of the Woods-Saxon potential are designed approximately for 
the nucleus $^{65}_{26}$Fe$_{39}$, of which $S_n$ = 4.18 MeV.
The diffuseness, the radius and the depth of the Woods-Saxon potential are 
0.67 fm, 5.08 fm, and $-$43.0 MeV, respectively.
Some one-particle levels possibly occupied by the N $\approx$ 40th 
neutron for some prolate deformation 
are denoted by the asymptotic quantum numbers, 
[$N n_z \Lambda \Omega$].
Positive-parity levels are plotted by solid curves.
The neutron numbers 40 and 50, which are obtained by filling in all 
lower-lying levels, are indicated with circles. 
See the text for details.}
\label{fig6}
\end{center}

\end{figure}


\vspace{2cm}
\begin{thebibliography}{99}
\bibitem{RKP69} Klapisch R, Thibault-Philippe C, Detraz C, Chaumont J, Bernas R
and Beck E 1969  {\it Phys. Rev. Lett.} {\bf 23} 652
\bibitem{CT75} Thibault C, Klapisch R, Rigaud C, Poskanzer A M, Prieels R,
Lessard L and Reisdorf W 1975 {\it Phys. Rev.} C {\bf 12} 644
\bibitem{EKW90} Warburton E K, Becker J A and Brown B A 1990 {\it Phys. Rev.} C 
{\bf 41} 1147
\bibitem{BM69} Bohr A and Mottelson B R 1969 {\it Nuclear Structure} vol 1 
(Massachusetts: Benjamin)
\bibitem{HM03} Hamamoto I and Mottelson B R 2003 {\it C. R. Phys.} {\bf 4} 433 
\bibitem{IH07} Hamamoto I 2007 {\it Eur. Phys. J. Special Topics} {\bf 150} 123 
\bibitem{HI09} Suzuki D {\it et al} 2009 {\it Phys. Rev. Lett.} {\bf 103} 152503 
\bibitem{BM75} Bohr A and Mottelson B R 1975 {\it Nuclear Structure}  vol 2
(Massachusetts: Benjamin)
\bibitem{IH07a} Hamamoto I 2007 {\it Phys. Rev.} C {\bf 76} 054319
\bibitem{IH04} Hamamoto I 2004 {\it Phys. Rev.} C {\bf 69} 041306(R)
\bibitem{IH05} Hamamoto I 2005 {\it Phys. Rev.} C {\bf 72} 024301 
\bibitem{IH06} Hamamoto I 2006 {\it Phys. Rev.} C {\bf 73} 064308 
\bibitem{RGN66} For example, see; Newton R G 1966 
{\it Scattering Theory of Waves and Particles} (New York: McGraw-Hill) 
\bibitem{BR64} Bodenstedt E and Rogers J D 1964 {\it in Perturbed Angular
Correlations} (Amsterdam: North-Holland) eds Karlsson E, Matthias E and 
Siegbahn K 
\bibitem{KA02} Asahi K {\it et al} 2002 {\it Nucl. Phys.} {\bf A704} 88c 
\bibitem{HU04} Ueno H {\it et al} 2004 {\it Nucl. Phys.} {\bf A738} 211
\bibitem{WG99} Geithner W 1999 {\it Phys. Rev. Lett.} {\bf 83} 3792 
\bibitem{GN05} Neyens G {\it et al} 2005 {\it Phys. Rev. Lett.} {\bf 94} 022501
\bibitem{VT08} Tripathi V {\it et al} 2008 {\it Phys. Rev. Lett.} {\bf 101} 
142504
\bibitem{DY07} Yordanov D T {\it et al} 2007 {\it Phys. Rev. Lett.} {\bf 99} 
212501
\bibitem{BB07} Bastin B {\it et al} 2007 {\it Phys. Rev. Lett.} {\bf 99} 022503
\bibitem{TG97} Glasmacher T {\it et al} 1997 {\it Phys. Lett.} {\bf B395} 163
\bibitem{BJ07} Jurado B {\it et al} 2007 {\it Phys. Lett.} {\bf B649} 43
\bibitem{LG09} Gaudefroy L {\it et al} 2009 {\it Phys. Rev. Lett.} {\bf 102} 
092501
\bibitem{PA08} Adrich P {\it et al} 2008 {\it Phys. Rev.} C {\bf 77} 054306
\bibitem{MS03} Sawicka M {\it et al} 2003 {\it Eur. Phys. J.} A {\bf 16} 51
\bibitem{AO00} Ozawa A, Kobayashi T, Suzuki T, Yoshida K and Tanihata I 2000 
{\it Phys. Rev. Lett.} {\bf 84} 5493
\bibitem{HS07} Hamamoto I and Shimoura S 2007 {\it J. of Phys.} {\bf G34} 2715
\bibitem{TO05} For example, see; Otsuka T, Suzuki T, Fujimoto R, Grawe H, 
and Akakishi Y 2005 {\it Phys. Rev. Lett.} {\bf 95} 232502

\end{thebibliography}
\end{document}